\documentclass[10pt,preprint]{aastex} 

\newcommand{\con}{\mathrm{const}}
\newcommand{\dif}{\mathrm{d}}
\newcommand{\me}{m_{\rm e}}

\newcommand{\mec}{m_{\rm e} c}

\newcommand{\mecc}{m_{\rm e} c^2}

\newcommand{\sigmaT}{\sigma_{\rm T}}
\newcommand{\alphae}{\alpha_{\rm e}}
\newcommand{\Aw}{A_{\rm w}}
\newcommand{\betaw}{\beta_{\rm w}}
\newcommand{\vw}{v_{\rm w}}
\newcommand{\dL}{d_{\rm L}}
\newcommand{\Dg}{D_\gamma}
\newcommand{\Dw}{D_{\rm w}}
\newcommand{\Dp}{D_p}

\newcommand{\Dmup}{D_{\mu p}}
\newcommand{\Dmumu}{D_{\mu \mu}}
\newcommand{\etaesc}{\eta_{\rm esc}}
\newcommand{\etasyn}{\eta_{\rm syn}}
\newcommand{\etadam}{\eta_{\rm dam}}
\newcommand{\etarad}{\eta_{\rm rad}}
\newcommand{\Fe}{F_{\rm e}}
\newcommand{\Fw}{F_{\rm w}}
\newcommand{\gammainj}{\gamma_{\rm inj}}
\newcommand{\gammasyn}{\gamma_{\rm syn}}

\newcommand{\gammadotrad}{\dot{\gamma}_{\rm rad}}
\newcommand{\gammadotsyn}{\dot{\gamma}_{\rm syn}}
\newcommand{\gammadotIC}{\dot{\gamma}_{\rm IC}}
\newcommand{\Gammae}{\Gamma_{\rm e}}
\newcommand{\Gammaw}{\Gamma_{\rm w}}
\newcommand{\Ie}{I_{\rm e}}
\newcommand{\Iw}{I_{\rm w}}
\newcommand{\kres}{k_{\rm res}}

\newcommand{\kinj}{k_{\rm inj}}
\newcommand{\Omegag}{\Omega_{\rm g}}
\newcommand{\Qinje}{Q_{\rm inj,e}}
\newcommand{\Qinjw}{Q_{\rm inj,w}}
\newcommand{\nele}{n_{\rm e}}
\newcommand{\nph}{n_{\rm ph}}
\newcommand{\nuFnu}{\nu F_\nu}
\newcommand{\gammares}{\gamma_{\rm res}}
\newcommand{\qe}{q_{\rm e}}
\newcommand{\rg}{r_{\rm g}}
\newcommand{\tacc}{t_{\rm acc}}
\newcommand{\tsyn}{t_{\rm syn}}
\newcommand{\trad}{t_{\rm rad}}
\newcommand{\tesce}{t_{\rm esc,e}}
\newcommand{\tescph}{t_{\rm esc,ph}}
\newcommand{\tdam}{t_{\rm dam}}
\newcommand{\tc}{t_{\rm c}}
\newcommand{\tcas}{t_{\rm cas}}
\newcommand{\uB}{u_{\rm B}}
\newcommand{\WB}{W_{\rm B}}
\usepackage{amsmath,amssymb}
\usepackage{bm}
\usepackage{cases}
\shorttitle{Gyroresonant Acceleration in Blazars}
\shortauthors{J. Kakuwa}

\begin{document}


\title{Stochastic Gyroresonant Acceleration for Hard Electron Spectra of Blazars: Effect of Damping of Cascading Turbulence}


\author{Jun Kakuwa\altaffilmark{1}}
\email{junkakuwa@hiroshima-u.ac.jp}


\altaffiltext{1}{Department of Physical Science, Hiroshima University, Higashi-Hiroshima, 739-8526, Japan}


\begin{abstract}
Stochastic acceleration of nonthermal electrons is investigated in the context of hard photon spectra of blazars. 
It is well known that this acceleration mechanism can produce a hard electron spectrum of $m \equiv \partial \ln \nele(\gamma)/\partial \ln \gamma = 2$ with the high-energy cutoff, called an ultrarelativistic Maxwellian-like distribution, where $\nele(\gamma)$ is an electron energy spectrum. 
We revisit the formation of this characteristic spectrum, considering a particular situation where the electrons are accelerated through gyroresonant interaction with magnetohydrodynamic wave turbulence driven by the turbulent cascade. 
By solving kinetic equations of the turbulent fields, electrons, and photons emitted via the synchrotron self-Compton (SSC) process, we demonstrate that in the non-test-particle treatment, the formation of a Maxwellian-like distribution is prevented by the damping effect on the turbulent fields due to the electron acceleration, at least unless an extreme parameter value is chosen. 
Instead, a softer electron spectrum with the index of $m \approx -1$ is produced if the Kolmogorov-type cascade is assumed. 
The SSC spectrum that originates from the resultant softer electron spectrum is still hard, but somewhat softer and broader than the case of $m=2$. 
This change of achievable hardness should be noted when this basic particle acceleration scenario is tested with observations of hard photon spectra accurately. 
\end{abstract}



\section{INTRODUCTION}
\label{sec:int}

It has been understood that an energy dissipation region in the subparsec-scale jets of a subset of the active galactic nucleus is observed in blazars by virtue of the relativistic beaming effect \citep[e.g.,][]{Urry1995a, Abdo2010c, Marscher2009a, Beckmann2012a}. 
Observations of blazars have been carried out from the radio to gamma-ray bands and revealed that there exist in that region nonthermal, relativistic electrons distributed in the energy range of a few orders of magnitude \citep[e.g.,][]{Maraschi1994a, Inoue1996a, Kataoka1999a, Sikora2001a, Kino2002a, Krawczynski2004a, Kusunose2008a, Tagliaferri2008a, Aharonian2009a, Tavecchio2010a, abdo2011b, abdo2011a, Dermer2012a, Zhang2012a, Rani2013a}. 
These electrons mainly contribute to observed photon spectra, while how they are accelerated is unclear so far. 

Recently, a possibility has been discussed that electrons accelerated by stochastic acceleration (SA) are observed in some blazars \citep[e.g.,][]{Katarzynski2006a, Ushio2009a, Ushio2010a, Asano2014a, Kakuwa2015a}. 
SA is a mechanism that accelerates charged particles in turbulent fields and is described as diffusion of the particles in momentum space \citep[e.g.,][]{Hall1967a, Ptuskin1988a, Terasawa1989a, Schlickeiser1998a, Schlickeiser2002a, Ohira2013a, Kimura2015a}. 
It has been implied that on some assumptions the acceleration time of the highest-energy electrons in blazars may be too long to be interpreted by acceleration at a (mildly) relativistic shock \citep{Inoue1996a, Garson2010a, Kakuwa2015a}. 
Such slow acceleration increases the role of SA because of its slower nature. 
Also, the association between SA and the behavior of the synchrotron peak frequency, flux amplitude, and spectral shape around the peak has been noticed \citep{Katarzynski2006a, Giebels2007a, Tramacere2011a}. 

One of the features of SA is its ability to naturally produce hard electron spectra. 
Actually, recent application of SA to blazars has been concerned with hard electron and/or photon spectra \citep{Sauge2004a, Katarzynski2006b, Giebels2007a, Lefa2011b, Lefa2011a, Tramacere2011a, Zheng2011a, Asano2015a}. 
In particular, a very hard spectrum of relativistic electrons with the index 
\begin{align}
 \label{eq:m}
 m \equiv \frac{\partial \ln \nele(\gamma)}{\partial \ln \gamma} 
\end{align}
of $m = 2$ with the high-energy cutoff, which is called an ultrarelativistic Maxwellian-like distribution, 
\begin{align}
 \label{eq:pileup}
 \nele(\gamma) \propto \gamma^2 \exp \left[ - \frac{\etarad(\gamma)}{a} \right] \,, 
\end{align}
has often been applied to observations, where $\nele(\gamma)$ is the number density of the accelerated electrons per unit $\gamma$, $\gamma$ the Lorentz factor, $\etarad(\gamma) \equiv \tacc(\gamma)/\trad(\gamma)$ the radiative cooling efficiency, $\tacc(\gamma)$ the acceleration time, $\trad(\gamma)$ the radiative cooling time, and $a$ approximately constant of $\mathcal{O}(1)$. 
This near-monoenergetic spectrum is produced when the electrons accelerated by SA are confined in the acceleration site and then pile up around the characteristic energy where the acceleration and the synchrotron or inverse Compton (IC) cooling balance. 

The formation of the pileup distribution 
\footnote{
In this paper we refer to equation (\ref{eq:pileup}) also as the pileup distribution. As shown by \citet{Schlickeiser1984a} within the test-particle treatment, a different form of pileup distribution from that due to SA can also take place as a result of acceleration by multiple shock waves. 
}
in given turbulent fields (i.e., the test-particle treatment) was investigated in detail by \citet{Schlickeiser1984a, Schlickeiser1985a} for a particular case of the momentum-independent scattering rate \citep[][]{Fermi1949a, Parker1958a, Drury1987a}. 
The case of the momentum-dependent scattering rate has also been studied by, e.g., \citet{Bogdan1985a}, \citet{Borovsky1986a}, and \citet{Stawarz2008a}. 

Employing a non-test-particle formulation unlike these previous works, we shall revisit the formation of the pileup distribution. 
When applying SA to observations, one considers the situation where there is a net energy transfer from turbulent fields excited by external forces or by different species to emitting electrons. 
The energy of the turbulence is then more or less damped by the electrons. 
Blazar emission models employing SA, however, have not examined the damping effect, which can be taken into account only by performing a non-test-particle calculation.

The purpose of this paper is to demonstrate the damping effect on the pileup formation. 
Using a simplified formulation, we calculate spectra of the turbulent fields, accelerated electrons, and photons emitted by the electrons via the synchrotron self-Compton (SSC) process by solving respective kinetic equations including the damping effect. 
As the damping mechanism (i.e., electron acceleration mechanism), we consider gyroresonant interaction with small-amplitude, magnetohydrodynamic wave turbulence propagating parallel or antiparallel  to the mean magnetic field with the phase speed of the Alfv{\'e}n speed. 
The energy of the turbulence is assumed to be injected at a low wavenumber where the resonance does not occur and to be transferred to higher wavenumber by the Kolmogorov-type energy cascade. 

We do not discuss the influence of both the injection mechanism and detailed properties of the turbulence \citep[see, e.g.,][for informative discussions]{Eilek1984a, Dung1990b, Pohl2000a, Schlickeiser2000a, Schlickeiser2002b}. 
A model assumption that is relevant to these points but has more direct influence on our results is the energy transfer rate of the turbulent cascade in wavenumber space (or, in general, the energy injection rate at each wavenumber). 
The role of the assumed energy transfer rate is like that of a spectral index of the turbulence in test-particle models, in which the index is arbitrarily given and plays a crucial role \citep[e.g.,][]{Dermer1996a}. 

In Section \ref{sec:dam}, we estimate the efficiency of the damping for parameter values previously obtained with test-particle models. 
Kinetic equations of turbulent fields, electrons, and photons are described in Section \ref{sec:kin}, followed by Section \ref{sec:ste}, in which a steady-state solution of this coupled system is shown for the case of inefficient electron escape from the acceleration site (i.e., efficient confinement). 
We then show that inefficient escape, which is a required condition to establish the pileup distribution, is accompanied by efficient damping. 
As a result, electrons are prevented from piling up around the high-energy cutoff, so that a Maxwellian-like distribution is not established. 
Section \ref{sec:sum} is a summary. 

\section{DAMPING EFFICIENCY}
\label{sec:dam}

In this section, we show that under the assumptions mentioned in Section \ref{sec:int}, damping of turbulent fields by electron acceleration is estimated to be efficient for parameter values obtained in some previous works applying SA to blazars. 

We introduce a quantity indicating how the damping is effective, damping efficiency, as 
\begin{align}
 \label{eq:etadam1}
 \etadam = \etadam(k) \equiv \frac{\tcas(k)}{\tdam(k)} \, , 
\end{align}
where $k$ is the wavenumber, $\tcas = \tcas(k)$ the spectral energy transfer time of the turbulent fields (or simply the cascade time) at a wavenumber $k$, and $\tdam = \tdam(k)$ the damping time, which is the timescale that energy of the turbulent fields with wavenumber $k$ is consumed by electron acceleration. 
Hereafter, arguments of a function will often be omitted when they are clear or irrelevant. 
The energy of the turbulent fields at wavenumber $k$ is supposed to be supplied by the turbulent cascade from lower wavenumber. 
The damping efficiency higher than unity, $\etadam(k) > 1$, implies that a wavenumber spectrum of the turbulent fields is being affected by electron acceleration at wavenumber $k$. 

We assume that the cascade time is estimated as the characteristic time determined by only velocity fluctuations of the turbulence $\delta u_k$ and its wavenumber $k$: $\tcas \sim (k \delta u_k)^{-1}$, which leads to the Kolmogorov-type cascade \citep[e.g.,][]{Goldstein1995a, Zhou2004a}. 
Noting that the total energy density of the turbulent fields with wavenumber $k$ is $k W(k) \sim \rho \delta u_k^2$, we define the cascade time as 
\begin{align}
 \label{eq:tcas}
 \tcas = \frac{1}{k \betaw c}\sqrt{\frac{2 \uB}{k W(k)}} 
\end{align}
\citep[e.g.,][]{Miller1995a}, where $W(k)$ is the energy density of the turbulent fields per unit $k$, $\rho = 2 \uB/(\betaw c)^2$ the mass density of the background plasma, $\betaw = \vw/c$ the nonrelativistic Alfv{\'e}n speed in units of $c$, $c$ the speed of light, and $\uB = B^2/(8 \pi)$ the energy density of the mean magnetic field $B$. 
The inertial range spectrum formed by the pure cascade has a constant energy transfer rate in $k$-space, $k W(k)/\tcas = \con$, at an equilibrium state. 
In this case we have $W(k) \propto k^{-5/3}$ from $\tcas \propto k^{-3/2} \left( W(k) \right)^{-1/2}$ (equation [\ref{eq:tcas}]). 

The damping time $\tdam$ depends on electron acceleration mechanisms at work. 
Nonthermal electrons are assumed to be accelerated through gyroresonant interaction with weakly turbulent magnetohydrodynamic waves propagating parallel or antiparallel to the mean magnetic field with the phase speed of the Alfv{\'e}n speed. 
The time evolution of the ensemble average of electron phase-space density $F$ is described by the Fokker--Planck equation based on the quasilinear approximation \citep[][]{Hall1967a, Schlickeiser2002a}: 
\begin{align}
 \label{eq:FP}
 \frac{\partial F}{\partial t} - \Omegag \frac{\partial F}{\partial \phi} = \frac{1}{p^2} \frac{\partial}{\partial x_{\sigma}} \left[ p^2 D_{\sigma \alpha} \frac{\partial F}{\partial x_\alpha} \right] \, , 
\end{align}
where $\Omegag$ is the gyrofrequency (including the sign of the electron charge), $p$ the electron momentum, $D_{\sigma \alpha}$ the Fokker--Planck coefficients, and $x_\sigma$ denotes spherical coordinates in momentum space. 
We neglect spatial inhomogeneity for simplicity. 
Assuming small anisotropy, one obtains the momentum-diffusion equation of $f$, the isotropic part of $F$, as \citep[e.g.,][and references therein]{Schlickeiser1989a, Schlickeiser2002a} 
\begin{align}
 \label{eq:dfdt}
 \frac{\partial f}{\partial t} = \frac{1}{p^2}\frac{\partial}{\partial p} \left[ p^2 \Dp \frac{\partial f}{\partial p} \right]\, ,  
\end{align}
where $\Dp = \left< D_{pp} - \Dmup^2/\Dmumu \right>_\mu$ is the momentum-diffusion coefficient, and $\left< \right>_\mu$ denotes the pitch-angle averaging. 
We use an approximated form of the diffusion coefficient given by (see appendix \ref{ap:Dp} for a brief derivation)
\begin{align}
 \Dg = \Dg(\gamma) &\approx \frac{\Dp}{(\mec)^2} \nonumber \\ 
 &\sim \frac{\gamma^2 \betaw^2 c}{\uB \left( \rg(\gamma) \right)^2} \int_{\kres(\gamma)} \frac{\dif k }{k} \WB(k) \label{eq:Dg2} \\ 
 &\sim \frac{\gamma^2 \betaw^2 c \zeta(\kres)}{\rg(\gamma)} \, , \label{eq:Dg} 
\end{align}
where $\Dg$ is the diffusion coefficient in $\gamma$ space, $p = \gamma \me v$, $\me$ the electron mass, $v \approx c$ the electron speed, $\rg = \rg(\gamma) \approx \gamma \mecc/(e B)$ the gyroradius of electrons, $e$ the charge of an electron, $\zeta(k) \equiv k \WB(k)/\uB$ the relative amplitude of the turbulent magnetic fields, $\WB(k) \approx W(k)/2$ the magnetic component of $W(k)$, and $\kres = \kres(\gamma) \equiv \left( \rg(\gamma) \right)^{-1}$. 
If $\WB(k)$ becomes smaller for higher $k$, the lower end of the integration is more important than the upper end. 
The turbulent fields with wavenumber $k \sim \kres(\gamma)$ mainly contribute to $\Dg$ of electrons with Lorentz factor $\gamma$. 
The amplitude $\zeta(k)$ is less than 1 by definition since $B$ does not consist of only the the nonrandom magnetic field component; $\zeta(k)$ may have to be $\zeta(k) \lesssim 0.1$ in the wavenumber range relevant to the electron acceleration to meet the assumption of weak turbulence. 

In this section, we use the following estimate of the damping time: 
\begin{align}
 \label{eq:tdam}
 \tdam \sim \frac{k W(k)}{\dot{\mathcal{U}}_{\rm acc}} \, , 
\end{align}
where 
\begin{align}
 \dot{\mathcal{U}}_{\rm acc} = \frac{\left( \gammares(k) \right)^2 \me c^2 \nele(\gammares(k))}{\tacc(\gammares(k))}
\end{align}
is the energy transfer rate per unit volume from the turbulent fields to the electrons by the gyroresonant interaction, 
\begin{align}
 \tacc = \tacc(\gamma) = \frac{\gamma^2}{\Dg(\gamma)}
\end{align}
is the acceleration time of electrons with the Lorentz factor $\gamma$, and $\gammares$ is the electron Lorentz factor defined by $k = 1/\rg(\gammares)$ (i.e., $\gammares = \gammares(k) = e B/(\mecc k)$). 
The expression of $\tdam$ will be reintroduced in Section \ref{sec:kin} (equation [\ref{eq:Gamw}]). 
Note that the actual $\tdam$ can be shorter than the estimate of equation (\ref{eq:tdam}) if $\nele(\gamma)$ is harder than $m=0$ (like a Maxwellian-like distribution, $m=2$) because electrons with higher Lorentz factor than $\gamma=\gammares(k)$ mainly contribute to the damping in this case (equation [\ref{eq:Gamw2}]). 

From equations (\ref{eq:etadam1}), (\ref{eq:tcas}), and (\ref{eq:tdam}), we obtain 
\begin{align}
 \label{eq:etadam2}
 \etadam(\kres)
 \sim& \frac{4 \pi \betaw \gamma^2 \me c^2 \nele(\gamma)}{(\zeta(\kres))^{1/2} B^2} \nonumber \\
 \sim& \left( \frac{8 \pi \sigmaT \me^2 c^4}{3 e} \right)^{1/2} \frac{\gamma^3 \nele(\gamma)}{\left( \etasyn(\gamma) \right)^{1/2} \zeta(\kres) B^{3/2}} \, , 
\end{align}
where $\sigmaT$ is the Thomson cross section. 
In equation (\ref{eq:etadam2}) $\betaw$ is replaced with the synchrotron cooling efficiency, 
\begin{align}
 \label{eq:etasyn}
 \etasyn = \etasyn(\gamma) \equiv \frac{\tacc}{\tsyn} \sim \frac{\sigmaT}{6 \pi e} \frac{B \gamma^2}{\betaw^2 \zeta(\kres)} \, , 
\end{align}
for convenience of evaluating $\etadam$, where $\tsyn = \tsyn(\gamma) = 6 \pi \me c/(\sigmaT B^2 \gamma)$ is the synchrotron cooling time. 

Now, we can evaluate the damping efficiency $\etadam$ for some previous works with equation (\ref{eq:etadam2}). 
When emission from the electron acceleration site makes a dominant contribution to a blazar spectrum, $\nele(\gamma)$ and $B$ in this site are constrained by a spectral fitting. 
Some SA models assuming such a situation have been applied to observations (without taking into account the damping effect), and so the damping efficiency can be evaluated. 
When $\etasyn(\gamma)$ is unknown, we assume for the evaluation that the high-energy cutoff of $\nele(\gamma)$ obtained by the spectral fitting results from the competition between the acceleration and synchrotron cooling, i.e., $\etasyn(\gamma) \approx 1$ for the cutoff Lorentz factor; if the IC cooling is actually stronger than the synchrotron cooling, $\etadam(\kres)$ is underestimated. 
When $\zeta(\kres)$ is unknown, the upper limit of $\zeta(\kres)$ (i.e., $\sim 0.1$) is substituted to show the lower limit of $\etadam(\kres)$. 

Note that the estimate by equation (\ref{eq:etadam2}) assumes that $\nele(\gamma)$ is known. 
Although larger amplitude $\zeta(\kres)$ at a resonance scale gives smaller damping efficiency $\etadam(\kres)$ in this case, in general, when one considers a coupled system of $\nele(\gamma)$ and $W(k)$, larger amplitude of the turbulence may act to inject a larger number of electrons, accelerate them to higher energy, and confine them in the acceleration site for a longer time. 
As a result, larger turbulence amplitude can enhance the damping effect. 
This is the case of a simple non-test-particle model that we will employ in Section {\ref{sec:kin}}. 

Table \ref{tab:etadam} presents the results of the evaluation of $\etadam(k)$ at the resonant wavenumber $\kres(\gamma)$ of electrons with the high-energy cutoff Lorentz factor. 
We find $\etadam > 1$ for some previous applications of SA, that is, the damping can affect the formation of $W(k)$ and $\nele(\gamma)$, although they did not consider the damping effect \footnote{For reference 2 in Table \ref{tab:etadam}, $\etadam(\kres)$ is calculated with $\nele(\gamma)$ at the steady state shown in \citet{Katarzynski2006b}, while they reproduced observed photon spectra with evolving $\nele(\gamma)$. At the moment when their SA model fits to the observations before reaching the steady state, the values of $\etadam(\kres)$ are $\gtrsim 3$ and $\gtrsim 2$ for references 2$^{{\rm a}}$ and 2$^{{\rm b}}$, respectively. \label{foo:katarzynski2006b}}. 
This does not always mean that these previous tests of SA are invalid since in the test-particle models the mechanisms of both energy supply to the turbulence at each $k$ and the damping are not specified. 
Qualitatively speaking, if energy supply to the turbulent fields is more rapid than our assumption $(k \delta u_k)^{-1}$, the damping effect is weaker than equation (\ref{eq:etadam2}), and vice versa. 
Anyway, Table \ref{tab:etadam} shows the importance of performing tests of SA in blazars more self-consistently. 

We have not mentioned another type of models that assume that the electron acceleration site is basically a source region of nonthermal electrons and that emission from a more extended region is observed \citep[e.g.,][]{Sauge2004a, Giebels2007a}. 
If this is the case, it seems difficult to evaluate the importance of the damping in the same manner because less information on the acceleration site can be extracted with such models. 
However, note that when gyroresonant acceleration is considered, application of a Maxwellian-like electron spectrum (equation [\ref{eq:pileup}]) as the injection spectrum into the emission region requires an extreme parameter value for the acceleration site in order to avoid effective damping from the discussion in Section \ref{sec:ste}. 
Also note that when the damping is effective, the energy spectrum of electrons diffusing from the acceleration site into the emission region is softer than that of electrons in the acceleration site because higher-energy electrons escape more slowly than lower-energy electrons, as will be seen in Figure \ref{fig:time}.

Unlike the other works in Table \ref{tab:etadam}, \citet{Asano2014a} show $\etadam < 1$. 
They calculated SA in a jet flow and investigated effects of radial dependence of physical parameters on superposed photon spectra. 
It seems important to compare this model with more samples \citep{Asano2015a}. 

\section{KINETIC EQUATIONS}
\label{sec:kin}

Following Section \ref{sec:dam}, we investigate, in the rest of this paper, the damping effect on the formation of spectra of turbulent fields, electrons, and photons, particularly focusing on the pileup formation. 
Keeping the basic assumptions, equation (\ref{eq:tcas}) and (\ref{eq:Dg}), we adopt a simple formulation. 

A diffusion equation in wavenumber space has been used to consider energy injection, cascading, and damping in calculation of turbulence spectra for wave-particle systems: 
\begin{align} 
 \label{eq:W}
 \frac{\partial W(k)}{\partial t} + \frac{\partial}{\partial k} \left\{ - k^2 \Dw \frac{\partial}{\partial k} \left[ \frac{W(k)}{k^2} \right] \right\} = \Gammaw W(k) + \Iw(k) 
\end{align}
\citep[e.g.,][]{Eichler1979b, Zhou1990a, Miller1995a, Miller1996a, Brunetti2007a}, where $\Dw = \Dw(k) \equiv k^2/\tcas$, and the cascade time $\tcas$ is given by equation (\ref{eq:tcas}). 
The second term on the left-hand side ($\equiv \partial \Fw / \partial k$) represents the energy cascade and acts to establish the Kolmogorov spectrum $q = 5/3$, where 
\begin{align}
 q \equiv - \frac{\partial \ln W(k)}{\partial \ln k} \, . 
\end{align}
The damping effect is represented by $\Gammaw = \Gammaw(k)$, which will be described after introducing the electron kinetic equation. 
If the damping is effective, which is the case of interest, energy transport in $k$-space is not pure cascade, so that $W(k)$ is modified from the Kolmogorov spectrum. 
The energy injection into the turbulent fields is represented by $\Iw = \Iw(k) = \Qinjw \delta(k-\kinj)$, where $\Qinjw$ is the injection rate per unit volume, and $\delta(k)$ is the Dirac delta function. 
The injection wavenumber $\kinj$ is low enough not to resonate with electrons. 

The injection wavenumber $\kinj$ is just a parameter to determine the lower boundary of $k$. 
The amplitude of the turbulent fields is determined by $\Qinjw$. 
When $W(\kinj)$ reaches steady state, the balance $\Fw(\kinj) - \Qinjw = 0$ is achieved, where 
\begin{align}
 \Fw(k) = - k^2 \Dw \frac{\partial}{\partial k} \left[ \frac{W(k)}{k^2} \right] \, . 
\end{align} 
Thereby, noting $q=5/3$ in the wavenumber range with no damping such as around $\kinj$, that is, 
\begin{align}
 \label{eq:zet}
 \zeta(k) = \zeta_0 \left( \frac{k}{k_0} \right)^{-2/3} \, , 
\end{align}
we find the relation $\Qinjw = (22/3) \zeta_0^{3/2} c \betaw k_0 \uB$. 
Fixing $k_0 = 2 \pi/R$, where $R$ is the spatial size of the acceleration site, we regard $\zeta_0$ as a parameter instead of $\Qinjw$. 

Equation (\ref{eq:W}) describes the isotropic turbulent fields, which cascade omnidirectionally in wavenumber space. 
On the other hand, if we assume that the turbulence consists of pure parallel or antiparallel propagating waves, that is, zero energy cascade in perpendicular direction, the diffusion equation in parallel-wavenumber space is probably appropriate to describe the evolution of $W(k) $\citep[as adopted in][]{Miller1995a}. 
However, the difference of these two one-dimensional treatments perhaps does not matter for our results because it does not affect the self-similarity that we will see in Section \ref{sec:ste}. 
We adopt the form of equation (\ref{eq:W}), considering only parallel and antiparallel propagating waves as the acceleration agent (equation [\ref{eq:Dg}]). 
Strictly speaking, we have to take into account the three-dimensional turbulent cascade and the acceleration by and the damping of not only parallel or antiparallel propagating waves but also oblique waves \citep[e.g.,][]{Schlickeiser1998a}), although it is beyond the scope of this paper. 

Using $\nele(\gamma)/(\mec) \approx 4 \pi p^2 f$ instead of $f$, we calculate the electron energy spectrum by the following equation modified from the momentum-diffusion equation (equation [\ref{eq:dfdt}]): 
\begin{align} 
 \label{eq:ne}
 \frac{\partial \nele(\gamma)}{\partial t} + \frac{\partial}{\partial \gamma} \left\{ - \gamma^2 \Dg \frac{\partial}{\partial \gamma} \left[ \frac{\nele(\gamma)}{\gamma^2} \right] \right\} + \frac{\partial}{\partial \gamma} \left[ \gammadotrad \nele(\gamma) \right] = \Gammae \nele(\gamma) + \Ie(\gamma) \, . 
\end{align}
The second term on the left-hand side ($\equiv \partial \Fe/\partial \gamma$) represents SA, where $\Dg$ is given by equation (\ref{eq:Dg}). 
The coefficient of the third term on the left-hand side, which represents the systematic energy loss by radiation cooling, $\gammadotrad = \gammadotsyn + \gammadotIC$, is calculated consistently with the photon production rate in the photon kinetic equation, where $\gammadotsyn$ and $\gammadotIC$ are, respectively, the contribution from the synchrotron and IC emission. 
Only the photons emitted by the accelerated electrons are considered as target photons of IC scattering. 
Spatial escape of the electrons from the acceleration site is represented by $\Gammae = \Gammae(\gamma) = - (\tesce + R/c)^{-1}$, where $\tesce = \tesce(\gamma) = R^2/\kappa_\|$ is the escape time, and $\kappa_\| \sim c \rg(\gamma)/(9 \zeta(\kres))$ is the spatial diffusion coefficient along the mean magnetic field \citep{Longair1992a, Schlickeiser2002a}. 
The electron injection into the acceleration process is represented by $\Ie = \Ie(\gamma) = \Qinje \delta(\gamma - \gammainj)/(\gammainj \mec^2)$. 
The injection Lorentz factor $\gammainj$, which gives the lower boundary of $\gamma$ space, is fixed to $10$. 
(Although this value of $\gammainj$ may be physically unrealistic if Alfv{\'e}n turbulence in an electron--proton plasma is considered, it is not significant for our purpose to demonstrate the damping effect on the pileup formation.)
We control the electron energy injection rate per unit volume by introducing a dimensionless parameter $\alphae$: 
\begin{align}
 \label{eq:Qinje}
 \Qinje  = \alphae \Fw(\kres(\gammainj)) \, . 
\end{align}
The energy source of the electron injection is the cascading turbulence, and $\alphae$ is the efficiency of the energy transfer from the turbulent fields to the nonthermal electrons measured at $\gamma=\gammainj$. 

The choice of the functional form of the escape time $\tesce$ is not important because we will focus on hard electron spectra, which are expected when the escape is inefficient. 
Our choice of $\tesce$ seems to give its lower limit considering bending of the magnetic field. 
If one considers more realistic $\tesce$, the escape perhaps becomes more inefficient. 
Note that this results in strengthening the importance of the damping effect because inefficient escape leads to efficient damping, as will be explained in Section \ref{sec:ste}. 

The damping rate $\Gammaw$ is introduced to be consistent with the energy gain rate of the electrons \citep{Eilek1979a, Eilek1984a, Brunetti2007a}: 
\begin{align}
 \int \dif k \, \Gammaw W(k) = \int \dif \gamma \, \gamma \mec^2 \frac{\partial \Fe}{\partial \gamma} \, , 
\end{align}
where 
\begin{align}
 \Fe = - \gamma^2 \Dg \frac{\partial}{\partial \gamma} \left[ \frac{\nele(\gamma)}{\gamma^2} \right] \, . 
\end{align}
Performing a partial integration, from equation (\ref{eq:Dg2}) and $\rg(\gamma) \approx \gamma \mecc/(e B)$, we get 
\begin{align}
 \label{eq:Gamw2}
 \Gammaw(k) \sim \frac{4 \pi e^2 \betaw^2}{\mec k} \int_{\gammares(k)}^\infty \dif \gamma \, \gamma^2 \frac{\partial}{\partial \gamma} \left[ \frac{\nele(\gamma)}{\gamma^2} \right] \, . 
\end{align}
The turbulence with wavenumber $k$ is damped by electrons with $\gamma > \gammares(k)$. 
Assuming that $\nele(\gamma)$ is smaller for higher $\gamma$, we use 
\begin{align}
 \label{eq:Gamw}
 \Gammaw(k) \sim \left. \frac{4 \pi e^2 \betaw^2}{\mec k} \left( \gammares(k) \right)^3 \frac{\partial}{\partial \gamma}\left[ \frac{\nele(\gamma)}{\gamma^2} \right] \right|_{\gamma = \gammares(k)} \, . 
\end{align}
This simplification does not affect the results in this paper, and hence equation (\ref{eq:Gamw}) is sufficient for the current purpose. 
A somewhat refined expression of the damping time $\tdam$ compared to equation (\ref{eq:tdam}) is $| \Gammaw |^{-1}$. 

The energy spectrum of isotropic and homogeneous photons, $\nph(\epsilon)$, emitted by the nonthermal electrons is calculated by solving the photon kinetic equation identical to equation (3) in \citet{Kakuwa2015a}, where $\epsilon$ is the photon energy in units of $\mecc$, and $\nph(\epsilon)$ is the number density of the photons per unit $\epsilon$ \citep{Jones1968a, Blumenthal1970a, Rybicki1979a, Li2000a, Finke2008a}. 

\section{STEADY-STATE SOLUTION} 
\label{sec:ste}

There are five parameters to be set to calculate $W(k)$, $\nele(\gamma)$, and $\nph(\epsilon)$: $R$, $B$, $\zeta_0$, $\betaw$, and $\alphae$. 
We adopt $R=10^{16}$ cm, $B=0.5$ G, $\zeta_0=1$, $\betaw=0.05$, and $\alphae=0.1$ to demonstrate the effect of the damping on these spectra. 
This parameter set is an example that leads to an extremely hard electron spectrum, the pileup electron distribution $m = 2$, when one neglects the damping effect. 
We show that such a parameter set, however, asymptotically forms a softer spectrum $\gamma \nele(\gamma) \approx \con$ (i.e., $m=-1$) by the damping effect below the high-energy cutoff. 
At the initial time, we assume that $W(k)$ satisfies equation (\ref{eq:zet}) and that there are no accelerated electrons (i.e., $\nele(\gamma)=0$). 

If the damping is neglected (i.e., $\Gammaw = 0$), equation (\ref{eq:W}) produces a turbulence spectrum of equation (\ref{eq:zet}). 
Then, the characteristic Lorentz factor $\gammasyn$ defined by $\etasyn(\gammasyn) \equiv 1$ can be described as 
\begin{align}
 \label{eq:gamsyn}
 \gammasyn^\prime \sim \left( \frac{216 \pi^3 \me^2 c^4 e}{\sigmaT^3}  \right)^{1/4} \frac{k_0^{1/2} \zeta_0^{3/4} \betaw^{3/2}}{B^{5/4}} 
\end{align}
from equation (\ref{eq:etasyn}). 
Hereafter, primed quantities indicate that they are evaluated with the damping effect neglected. 
At $\gamma = \gammasyn^\prime$, the acceleration and synchrotron cooling balances and the high-energy cutoff is introduced for the electron spectrum. 

The condition required for the formation of the pileup distribution is $\etaesc^\prime(\gammasyn^\prime) \ll 1$ when the damping is neglected \citep{Schlickeiser1984a, Schlickeiser1985a, Bogdan1985a, Borovsky1986a, Stawarz2008a}, where 
\begin{align}
 \label{eq:etaesc}
 \etaesc = \etaesc(\gamma) &\equiv \frac{\tacc}{\tesce} \nonumber \\ 
 &\sim \left( \frac{\me c^2}{3 e} \right)^2 \frac{\gamma^2}{B^2 \betaw^2 R^2 \left(\zeta(\kres)\right)^2}
\end{align}
is the escape efficiency. 
We can calculate $\etaesc^\prime(\gamma)$ by just substituting equation (\ref{eq:zet}) into equation (\ref{eq:etaesc}). 
The adopted parameter set given at the beginning of this section is in the parameter region of inefficient escape: $\etaesc^\prime(\gammasyn^\prime) \sim 10^{-5} \ll 1$, where $\gammasyn^\prime \sim 4\times 10^4$. 
The first term on the right-hand side of equation (\ref{eq:ne}) is not important in our situation. 

\begin{figure}
\includegraphics{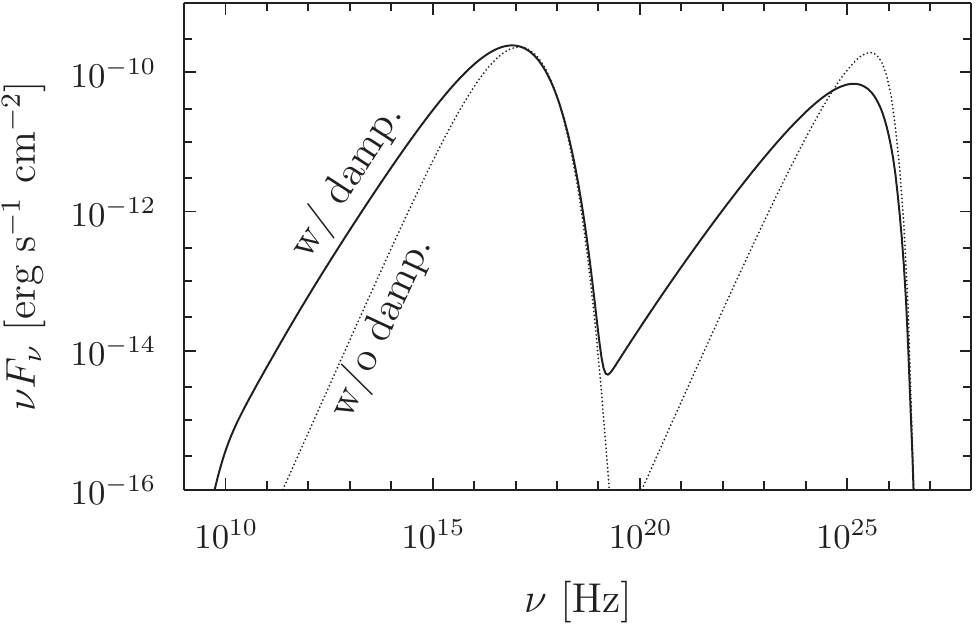}
\caption{
Solid line: steady-state SSC spectrum in the $\nuFnu$ representation. 
Damping of turbulence by electron acceleration is effective for the adopted parameter set: $R=10^{16}$ cm, $B=0.5$ G, $\zeta_0=1$, $\betaw=0.05$, $\alphae=0.1$, $\delta=10$, $z=0.031$, $\dL=4.14 \times 10^{26}$ cm, and $V=4 \pi R^3/3$. 
Dotted line: steady-state SSC spectrum in the case with the damping effect neglected. 
The same parameter values as the solid line are adopted, except that the flux amplitude and $\Qinje$ are adjusted for ease of comparison. }
\label{fig:SSC}
\end{figure}

The solid line in Figure \ref{fig:SSC} represents the photon spectrum obtained by solving the kinetic equations for $W(k)$, $\nele(\gamma)$, and $\nph(\epsilon)$ until the steady state. 
Approximating the acceleration site as a spherical, homogeneous blob moving with the Doppler factor $\delta$, we calculate a $\nuFnu$ form spectrum by
\begin{align}
 \label{eq:nuFnu} 
 \nuFnu = \frac{\delta^4}{4 \pi \dL^2} \frac{\epsilon^2 \mecc \nph(\epsilon) V}{\tescph} \, , 
\end{align}
where $\nu$ is the observed photon frequency given by $h \nu = \mecc \epsilon \delta / (1+z)$, $h$ the Planck constant, $z$ the redshift, $\dL$ the luminosity distance, $V$ the volume of the blob, and $\tescph = R/c$ the photon escape time. 
We choose the values of the additional parameters as $\delta=10$, $z=0.031$, $\dL=4.14 \times 10^{26}$ cm, and $V=4 \pi R^3/3$. 

In Figure \ref{fig:SSC}, for comparison a $\nuFnu$ spectrum calculated with the damping effect neglected is shown with the dotted line for the same parameters, except that the flux amplitude and $\Qinje$ are adjusted for ease of comparison. 
The spectrum originates from electrons of the pileup distribution, so that the hardest part of the $\nuFnu$ spectrum is proportional to $\nu^{4/3}$ for both synchrotron and IC spectrum. 
The solid line shows a softer and broader spectrum than the dotted line. 
This is because the damping works effectively (and prevents electrons from forming a spectrum harder than $m \approx -1$, as described later in this section). 
It is therefore important to take account of the damping effect when one interprets, with SA, a very hard photon spectrum that requires a near-monoenergetic $\nele(\gamma)$ like the pileup distribution. 

Instead of performing calculations for various parameter sets, here we see that in general, unless $\alphae$ is extremely small (i.e., a low electron injection rate), inefficient escape results in efficient damping; in other words, formation of the pileup electron distribution is affected by the damping. 
This can be easily proved by confirming that the damping is evaluated as {\it efficient} even when we start from the assumption that both the damping and escape are {\it inefficient}. 

First, suppose that the damping is inefficient. 
Then, we can replace $\kres(\gammainj)$ in equation (\ref{eq:Qinje}) with an arbitrary $k$ since the pure turbulent cascade leads to $\Fw = \con$ in the inertial range. 
From equation (\ref{eq:tdam}), $\tcas = 2 k/\Aw(k)$, equation (\ref{eq:Qinje}), and $\Fw(k) \sim \Aw(k) W(k)$, where we denote the systematic energy cascade rate per unit volume as $\Aw(k)$, equation (\ref{eq:etadam1}) is expressed by nondimensional quantities as 
\begin{align}
 \label{eq:etadam3}
 \etadam^\prime(\kres(\gamma)) \sim \frac{2 \alphae}{\eta_1^\prime(\gamma)} \frac{\gamma}{\gammainj} \, . 
\end{align}
Here, for convenience, $\eta_1 = \eta_1(\gamma)$ and $\tau_1 = \tau_1(\gamma)$ are respectively introduced by $\eta_1 \equiv \tacc/\tau_1$ and $\gamma \nele(\gamma) \equiv \Qinje \tau_1(\gamma)/(\gammainj \mec^2)$. 
Second, suppose that the escape is also inefficient (i.e., $\etaesc^\prime(\gammasyn^\prime) \ll 1$). 
The electrons then pile up around $\gammasyn^\prime$, that is, $\gamma \nele(\gamma)$ peaks around $\gammasyn^\prime$, and accordingly $\eta_1^\prime(\gamma)$ at $\gamma = \gammasyn^\prime$ becomes roughly equivalent to $\etaesc^\prime(\gammasyn^\prime)$. 
As a result, in the case of $\etaesc^\prime(\gammasyn^\prime) \ll 1$, $\gammainj \ll \gammasyn^\prime$, and a moderate value of $\alphae$ (like the adopted parameters), we have $\etadam^\prime(\kres(\gammasyn^\prime)) \gg 1$ from equation (\ref{eq:etadam3}). 
This conflicts with our first assumption of neglecting the damping. 
Therefore, efficient damping seems difficult to avoid when escape is inefficient. 
Equation (\ref{eq:etadam3}) states that efficient injection, efficient confinement (or inefficient escape), and efficient acceleration of electrons to higher energy lead to efficient damping. 
We should note that equation (\ref{eq:tdam}) that we used for equation (\ref{eq:etadam3}) probably overestimates the damping time as mentioned in Section \ref{sec:dam}. 

\begin{figure}
\includegraphics{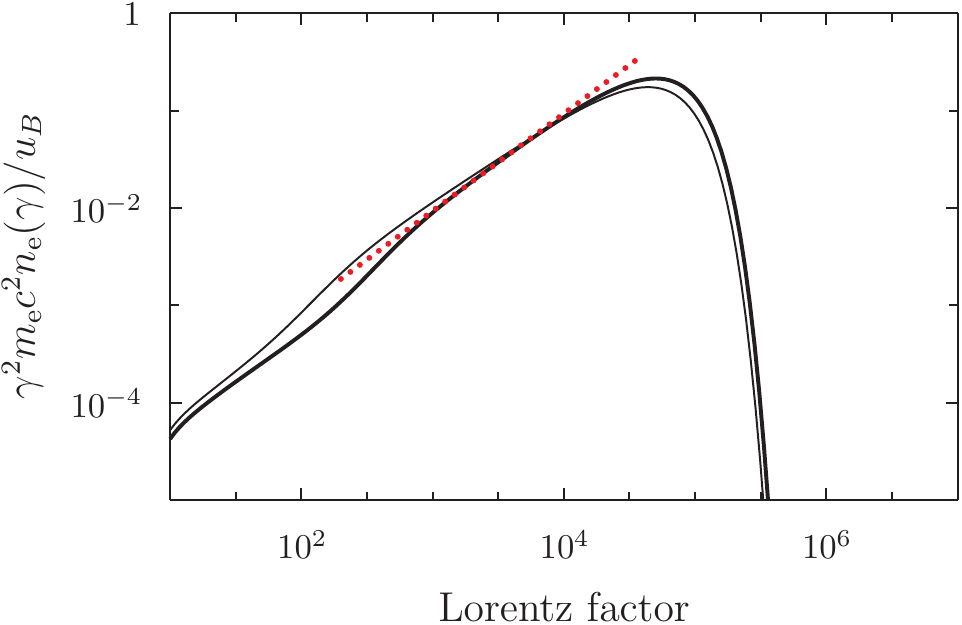}
\caption{Solid lines: steady-state energy spectra of accelerated electrons under the situation where turbulent fields are effectively damped owing to the electron acceleration. 
The thick line represents the case with only the synchrotron cooling, while the thin line represents the case with both the synchrotron and IC cooling. 
The adopted parameter set is the same as Figure \ref{fig:SSC}. 
Dotted line: asymptotic electron spectrum given in equation (\ref{eq:asy}), which is valid for the energy range where the balances are approximately established between the acceleration and radiation cooling of the electrons ($t_{\rm rad}/t_{\rm acc} \approx \con$) and between the energy cascade and damping of the turbulent fields ($t_{\rm cas}/t_{\rm dam} \approx \con$). 
The right end of this line is set at the Lorentz factor where the acceleration and cooling balance when the damping effect is neglected, i.e., $\gamma = \gammasyn^\prime$ (equation [\ref{eq:gamsyn}]). }
\label{fig:ne}
\end{figure}

When the bulk of electrons reaches $\gammasyn^\prime$ (equation [\ref{eq:gamsyn}]) before reaching the steady state, they form the cutoff by the competition between the acceleration and radiation cooling. 
The estimate of the cutoff Lorentz factor is not affected by the damping because the damping occurs resonantly (i.e., $\kres = \rg^{-1}$), that is, $q=5/3$ is kept for $k \lesssim \kres(\gammasyn^\prime)$. 
This means that the synchrotron cooling efficiency $\etasyn(\gamma)$ becomes higher $(\gtrsim 1)$ for higher $\gamma \, (\gtrsim \gammasyn^\prime)$ owing to $\etasyn \propto \gamma^{3-q}$ (equation [\ref{eq:etasyn}]). 
In this case, the cutoff is formed in $\nele(\gamma)$ to balance the positive, diffusive flux by SA with the negative, advective flux by the cooling in $\gamma$ space, i.e., zero electron flux in energy space $\Fe - | \gammadotrad | \nele(\gamma) = 0$. 
Figure \ref{fig:ne} shows the steady-state electron spectra, $\gamma^2 \nele(\gamma)$, obtained by solving the kinetic equations with and without the IC cooling with the thin and thick solid lines, respectively. 
One can see the cutoff at $\gamma = \gammasyn^\prime$, which is indicated by the right end of the red dotted line. 

At the time when electrons form the cutoff, a balance is established also in the lower-energy side of $\gammasyn^\prime$ by the same mechanism in the case of no damping, but not in the case of effective damping. 
Since $W(k)$ is decreased by the damping from the initial amplitude (equation [\ref{eq:zet}]) at $k \gtrsim \kres(\gammasyn^\prime)$ in the case of effective damping, the acceleration is slowed at $\gamma \lesssim \gammasyn^\prime$ by the lowered $W(k)$. 
Then, the cooling becomes efficient relative to the acceleration, so that the region where the cooling dominates for the systematic energy change of electrons extends also to the lower-energy side of $\gammasyn^\prime$, unlike the case without the damping. 
In this region, the system adjusts both $\nele(\gamma)$ and $W(k)$ (or the diffusion coefficient $\Dg(\gamma)$) to establish the balance $\Fe - | \gammadotrad | \nele(\gamma) = 0$ to reach the steady state. 
We refer to this energy/wavenumber range as the damping region for convenience. 

\begin{figure}
\includegraphics{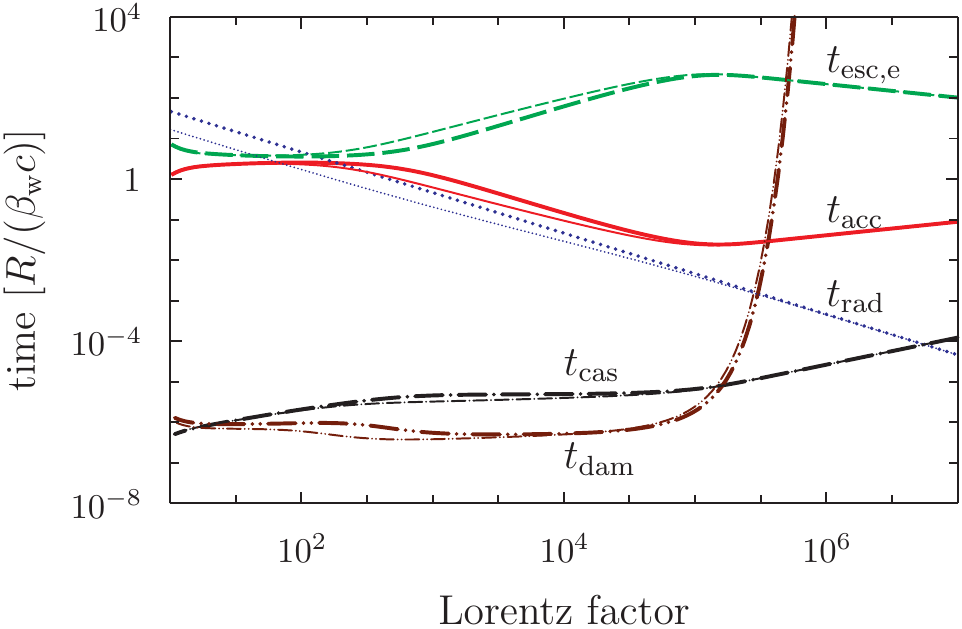}
\caption{Characteristic times of turbulent fields and accelerated electrons at the steady state. 
The horizontal axis represents the electron Lorentz factor $\gamma$. 
The timescales for the turbulence with wavenumber $k$ are plotted at $\gamma = \gammares(k)$. 
The thick lines represent the case taking into account only the synchrotron cooling, while the thin lines represent the case taking into account both the synchrotron and IC cooling. 
Lines show cascade time (black dot-dashed lines), damping time (brown double-dot-dashed lines), acceleration time (red solid lines), escape time (green dashed lines), and radiation cooling time (blue dotted lines). }
\label{fig:time}
\end{figure}

In Figure \ref{fig:time}, we show the characteristic times of the turbulent fields and the electrons at the steady state. 
The damping region is seen in the range of $10^2 \lesssim \gamma \lesssim \gammasyn^\prime (\sim 4\times 10^4)$. 
As shown in this figure, the balances between the acceleration and synchrotron/IC cooling and between the cascading and damping are established in the form of $\etasyn = \tacc/\tsyn \approx \con$ and  $\etadam = \tcas/\tdam \approx \con$ in this region, respectively. 
Note that turbulent fields with $q>2$ formed by the effective damping accelerate higher-energy electrons more rapidly than lower-energy electrons since $\tacc \propto \gamma^{2-q}$, so that the acceleration and cooling can balance in such a way (i.e., $\etasyn \approx \con$). 
This equilibrium state is a particular case of $a=0$ in equation (\ref{eq:pileup}).

We can find the asymptotic solution of the damping region at the steady state simply as follows. 
Let us start from $\etarad = \con$, where $\etarad \equiv \tacc/\trad = \gamma | \gammadotrad |/\Dg$, $\trad = \gamma/| \gammadotrad |$, and $\gammadotrad \propto \gamma^r$ (e.g., $r=2$ for the synchrotron emission). 
Then, $q$ and $r$ are related by $q=1+r$ owing to $\Dg \propto \gamma^q$. 
A constant $r$ leads to a constant $q$. 
Noting $\Gammaw = - \etadam/\tcas$, we obtain $\etadam = (q+2)(3q-5)/2 = (r+3)(3r-2)/2$ from equation (\ref{eq:W}) with $\partial W(k)/\partial t = 0$ and $\Iw = 0$. 
Now, $\etadam$ is also constant. 
Using $- \Gammaw = \etadam/\tcas$, $\tcas \propto k^{(q-3)/2}$ from equation (\ref{eq:tcas}) and $\Gammaw \propto k^{-1-m}$ from equation (\ref{eq:Gamw}), where $m \equiv \partial \ln \nele(\gamma)/\partial \ln \gamma$, we obtain $m = (q-5)/2 = (r-4)/2$. 
Finally, imposing the balance $\Fe - | \gammadotrad | \nele(\gamma) = 0$, where $\gammadotrad = - \Dg \etarad/\gamma$, we obtain $\etarad = 2-m = 4-r/2$. 
Substituting $r=2$ as an example, we get 
\begin{align}
 \label{eq:q}
 \left\{
 \begin{array}{lll}
  q &= 1 + r   &= 3 \\
  m &= r/2 - 2 &= -1 \\
  \etadam &= (r+3)(3r-2)/2 &= 10 \\
  \etarad &= 4 - r/2       &= 3 \, . 
 \end{array}
 \right.
\end{align}
When the damping is effective, the electron spectrum approaches to a non-monoenergetic spectrum, $\gamma \nele(\gamma) \approx \con$, when $r \approx 2$. 
The balance required to form a Maxwellian-like distribution, $\Fe = 0$, is not realized around the cutoff even if the electron escape is inefficient since systematic energy change of electrons is dominated by radiation cooling. 

Strength of the $\gamma$ or $k$ dependence of the electron energy-loss rate, that of the acceleration rate, and that of the spectral energy transfer rate are simply related to each other to reach the equilibrium. 
For example, in Figure \ref{fig:ne} we can see softening of an electron spectrum by the IC energy loss with effective Klein--Nishina suppression. 

From $\gamma \approx e B/(\mecc \kres)$ and equations (\ref{eq:etasyn}), (\ref{eq:etadam1}), (\ref{eq:tcas}), and (\ref{eq:Gamw}), we can write $\zeta(k)$ and $\nele(\gamma)$ with $B$ and $\betaw$ as
\begin{align}
 \label{eq:asy}
 \left\{
 \begin{aligned}
  k^2 \zeta(k)         &= \frac{\sigmaT e}{6 \pi \me^2 c^4} \frac{F(\gammares)}{\etarad} \frac{B^3}{\betaw^2} \\ 
  \gamma \nele(\gamma) &= \left( \frac{\sigmaT}{96 \pi^3 e \me^2 c^4} \right)^{1/2} \frac{F(\gamma)^{1/2} \etadam}{|m-2| \etarad^{1/2}} \frac{B^{5/2}}{\betaw^2} \, , 
 \end{aligned}
 \right.
\end{align}
where $F(\gamma)$ is defined by $\etarad \equiv F(\gamma) \etasyn(\gamma)$ (i.e., $\gammadotrad = F(\gamma) \gammadotsyn$). 
Since $\etarad$, $\etadam$, and $m$ are known from equation (\ref{eq:q}), equation (\ref{eq:asy}) gives the asymptotic spectra in the damping region at the steady state. 
The left-hand side of equation (\ref{eq:asy}) is approximately constant, though an uncertain parameter coming from the SSC process is included. 
In Figure \ref{fig:ne}, this analytical form of the asymptotic electron spectrum is shown by the red dotted line with only the synchrotron cooling considered. 

It may (or may not) seem strange that the cutoff $\gamma$ is introduced at $\gammasyn^\prime \sim 4\times 10^4$ (not around $10^2$) when one sees Figure \ref{fig:time}. 
In this regard, note that the cutoff region $\gamma \gtrsim \gammasyn^\prime$ balances (i.e., $\Fe - | \gammadotrad | \nele(\gamma) = 0$) before the formation of the damping region, while Figure \ref{fig:time} displays the steady state. 
The cutoff is introduced by the radiation cooling, and its position can be estimated as $\gammasyn^\prime$ (equation [\ref{eq:gamsyn}]) by the initial parameters. 
This usual cutoff does not emerge when one treats the situation of the efficient damping without solving the spectrum of turbulent fields \citep[e.g.,][]{Eilek1984a, Ohno2002a}. 

The extension of the damping region to lower energy/higher wavenumber can be suppressed by the escape effect. 
In the lower-energy side of the damping region, $\nele(\gamma)$ is determined by the acceleration and escape effect. 
We do not focus on spectra in this region because more exact formulation seems to be needed. 

The electron index $m$ in the damping region depends on the energy supply mechanism to the turbulent fields. 
Even within the framework of the turbulent cascade, other values of $m$ are expected by different $k$ dependence of the cascade time $\tcas(k)$. 
For example, if we may consider the Kraichnan-type cascade $\tcas \propto k^{-2} \left( W(k) \right)^{-1}$ \citep[e.g.,][]{Dobrowolny1980a, Goldstein1995a, Zhou2004a, Brunetti2007a} instead of the Kolmogorov one in the derivation of equation (\ref{eq:q}), we obtain a harder spectrum with the index of $m = 0$ when $r=2$. 
This is because the turbulent energy is transferred more slowly at higher wave number under efficient damping, unlike the Kolmogorov case, and accordingly, fewer lower-energy electrons relative to higher-energy electrons are needed to establish the equilibrium. 

\section{SUMMARY}
\label{sec:sum}

SA of nonthermal electrons, which is caused by damping of turbulent fields, has been discussed in blazars, in particular, as the origin of very hard photon/electron spectra (Section \ref{sec:int}). 
Test-particle models employing SA have been discussed in many papers, while models explicitly including the damping effect have not been considered so far. 
In this paper, specifying the damping mechanism, we have investigated the influence of the damping effect particularly on the formation of a Maxwellian-like spectrum (equation [\ref{eq:pileup}]), which is a well-known very hard electron spectrum produced by SA (sometimes called the pileup distribution). 
As the damping mechanism, we have assumed gyroresonant interaction with small-amplitude, magnetohydrodynamic wave turbulence propagating parallel to the mean magnetic field with the phase speed of the Alfv{\'e}n speed. 
This situation has often been mentioned in the literature. 

We have shown in Section \ref{sec:dam} and Table \ref{tab:etadam} that the damping is estimated to be efficient when we adopt parameter values previously obtained with test-particle models. 
This implies the importance of taking into account the damping effect when one examines the applicability of SA to blazars. 

Solving kinetic equations of the turbulent fields, electrons, and photons described in Section \ref{sec:kin}, we have shown in Section \ref{sec:ste} that inefficient electron escape from the acceleration site, which is a required condition to establish a Maxwellian-like spectrum, is accompanied by efficient damping (equation [\ref{eq:etadam3}]). 
As a result, electrons are prevented from piling up around the high-energy cutoff, so that a Maxwellian-like electron spectrum is not established. 
At the equilibrium state, the ratio of the radiation cooling time to the acceleration time becomes independent of $\gamma$ owing to a steepened turbulent spectrum by the damping, so that zero electron flux in $\gamma$-space is realized. 
The efficient damping leads to a softer electron spectrum, $\gamma \nele(\gamma) \approx \con$ (Figure \ref{fig:ne}), than the case of a Maxwellian-like spectrum ($m=2$). 
We have also presented a simple analytical form of the asymptotic spectra of the electrons and turbulent fields at the steady state under efficient damping (equation [\ref{eq:asy}]), though they have an uncertain parameter coming from the SSC process. 

Unlike the change of achievable hardness of $\nele(\gamma)$ from the electron index of $m=2$ to $m\approx-1$ due to the damping effect, the electron maximum energy is limited by the competition between the acceleration and radiation cooling as in the case without the damping. 
This usual cutoff does not emerge when one treats the situation of efficient damping without solving the evolution of turbulent fields \citep[][]{Eilek1984a, Ohno2002a}. 

The SSC spectrum under efficient damping is still hard, but somewhat softer and broader than the case of no damping because of the change of the hardness of $\nele(\gamma)$. 
This effect should be noted when this basic particle acceleration scenario is compared with observations of hard blazar spectra accurately. 
If the necessity of electron spectra harder than $m = -1$ is observationally confirmed, the situation examined in this paper is insufficient to explain them, and we need to consider different situations.

\acknowledgments
The author is deeply grateful to K. Toma for his valuable comments and advice during the course of this work and to the anonymous referee for his/her useful suggestions. 
The author would also like to thank R. Yamazaki, T. Kato, Y. Kojima, K. Yamamoto, and N. Okabe for their valuable advice.

\appendix
\section{A brief derivation of the momentum-diffusion coefficient}
\label{ap:Dp}

The pitch-angle-averaged momentum-diffusion coefficient due to gyroresonant interaction between parallel-propagating transverse magnetohydrodynamic waves and charged particles can be estimated by the following procedure. 
The purpose here is to present a simple calculation of the coefficient that is identical to the rigorous expression in the approximation of equation (\ref{eq:fast}). 
The resultant expression is used in equation (\ref{eq:Dg}) (see \citet{Schlickeiser2002a} and references therein for a full derivation of all Fokker--Planck coefficients). 

The gyroresonance condition between the above-mentioned waves and a charged particle is given by $\omega - k_\| v_\| = \Omegag$, where $\omega = \pm \vw k_\|$ is the wave frequency, $\vw$ the phase speed parallel to the mean magnetic field $\bm{B}$ ($\left| \bm{B} \right| \equiv B$), $k_\|$ the wavevector component parallel to $\bm{B}$, $v_\|$ the velocity of the particle parallel to $\bm{B}$, and $\Omegag$ the gyrofrequency (including the sign of charge). 
The positive and negative signs in the dispersion relation indicate the parallel ($\omega/k_\| > 0$) and antiparallel ($\omega/k_\| < 0$) propagation to $\bm{B}$, respectively. 
Here, positive (negative) $\omega$ denotes circular polarization in the direction of gyrorotation of positive (negative) charges. 
A particle $(v, \mu)$ can resonate with waves $(\omega, k_\|)$ satisfying the resonance condition, where $v$ is the particle speed and $\mu = v_\|/v$ is the pitch-angle cosine. 
The resonance can occur with waves propagating in both directions. 

We take into consideration particles with 
\begin{align}
 \label{eq:fast}
 v > \left| v_\| \right| \gg \vw 
\end{align}
(i.e., $\left| \mu \right| \gg \vw/v$) for simplicity. 
The resonance condition can be rewritten as 
\begin{align}
 \label{eq:kres}
 k_\| \approx - \Omegag/v_\| \, . 
\end{align}
The resonant parallel wavenumber is then approximately independent of the propagation direction. 
The lowest $\left| k_\| \right|$ in resonance is $\left| \Omegag \right| / v = \rg^{-1}$, where  $\rg \equiv v/\left| \Omegag \right|$ is the gyroradius (for $\mu=0$). 

Consider scattering by waves in a wavenumber range $\Delta k_\|$ around the wavenumber $k_\|$ exactly satisfying the resonance condition. 
Suppose that a particle and these waves are coherent in phase for the time $\Delta \tc$ in the plasma rest frame (hereafter, the lab frame). 
The wave magnetic field scatters the particle, conserving the particle energy in the frame moving with the parallel wave phase velocity (called the wave frame), in which the wave electric field vanishes. 
Noting that in the guiding center frame the frequency of the waves in resonance is nearly equal to the gyrofrequency and that the waves are approximately stationary for the particle in the sense of equation (\ref{eq:fast}), we can make a rough estimate of the change rate of the relative phase between the waves and the particle as $\left| v_\| \right| \Delta k_\|$ in the lab frame. 
Hence, we have \citep{Wentzel1974a, Kulsrud2005a} 
\begin{align}
 \label{eq:Deltc}
 \Delta \tc \approx \frac{2 \pi}{\left| v_\| \right| \Delta k_\|} \, .
\end{align}

The change in particle momentum $\Delta p$ associated with that in a pitch-angle cosine due to the scattering, $\Delta \mu = \dot{\mu} \Delta \tc$, where $\dot{\mu}$ is the change rate of the pitch-angle cosine, is given by \citep{Ostrowski1997a}
\begin{align}
 \label{eq:Delp}
 \Delta p \approx \pm p \left( \frac{\vw}{v} \right) \Delta \mu
\end{align}
by performing Lorentz transformations between the lab frame and the wave frame under the assumption of equation (\ref{eq:fast}), where both $\Delta p$ and $\Delta \mu$ are measured in the lab frame, and the positive and negative signs again indicate the propagation direction parallel and antiparallel to $\bm{B}$, respectively. 
Since the momentum change is a much slower process than the pitch-angle change due to equation (\ref{eq:fast}), we may neglect the momentum change during the coherence time $\Delta \tc$ to calculate $\dot{\mu}$. 
The rate $\dot{\mu}$ along the unperturbed helical trajectory is obtained by $\dot{\mu} \approx \dot{p}_\|/p \approx \qe \left( \bm{v} \times \delta \bm{B}_\perp \right)_\| / (p c) = \Omegag \sqrt{1-\mu^2} (\delta B_\perp / B) \sin \phi(t)$, where $t$ is the time, $p_\| = \mu p$, $p$ is the momentum, $\qe$ the particle charge, $\bm{v}$ the particle velocity, $\delta B_\perp = \left| \delta \bm{B}_\perp \right|$ the magnetic field amplitude of the resonant waves around $(\omega, k_\|)$, and $\phi(t) = (k_\| v_\| - \omega + \Omegag)t + \con$ \citep[e.g.,][]{Blandford1987a, Kulsrud2005a, Blasi2013a}. 
We used $B$ to normalize $\delta B_\perp$. 
The farther the waves and the particle are from the resonance, the smaller $\left| \Delta \mu \right|$ becomes. 
Since the resonance is expected to occur at a random wave phase (that is, the constant term of $\phi(t)$ is random), the sign of $\Delta \mu$ is also random. 
On average, we get $\dot{\mu} \approx 0$ and 
\begin{align}
 \label{eq:dotmu}
 \dot{\mu}^2 \approx \frac{1}{2} \Omegag^2 (1-\mu^2) \left( \frac{\delta B_\perp}{B} \right)^2 \, . 
\end{align}
The resonant waves propagating in each direction can scatter the particle in pitch angle. 
The contribution to $\dot{\mu}^2$ by the waves in each direction ($\equiv \dot{\mu}_\pm^2$) depends on their amplitude ($\equiv \delta B_\pm$). 
The pitch-angle diffusion coefficient is hence written as $D_\mu = D_{\mu+} + D_{\mu-}$, where 
\begin{align}
 \label{eq:Dmu}
 D_{\mu\pm} \approx \frac{1}{2} \dot{\mu}_\pm^2 \Delta \tc \approx \frac{\pi}{2} \left| \Omegag \right| \frac{1-\mu^2}{\left| \mu \right|} \frac{B_\pm^2 (k_\|)}{\rg B^2} 
\end{align}
\citep[e.g.,][]{Blandford1987a, Schlickeiser1989a} from equation (\ref{eq:Deltc}) and (\ref{eq:dotmu}), and $B_\pm^2 (k_\|) = (\delta B_\pm)^2/\Delta k_\|$ is the wave power spectrum \footnote{To be precise, the resonance width $\Delta k_\|$ also depends on the wave amplitude and hence on the propagation direction.} at the resonant wavenumber given by equation (\ref{eq:kres}), which is approximately the same for the resonant waves in both directions. 
If the waves are monochromatic, the resonant particle just oscillate in pitch angle and the pitch-angle diffusion does not occur. 

Unlike the pitch-angle diffusion, momentum diffusion needs scattering by waves propagating in both directions. 
Clearly, if resonant waves propagate unidirectionally, energy of scattered particles is kept constant in the relevant wave frame, so that the momentum diffusion, which leads to stochastic acceleration, does not occur. 
The momentum diffusion results from combination of small momentum changes due to $D_{p+}$ and $D_{p-}$, where $D_{p \pm} \equiv p^2 (\vw/v)^2 D_{\mu \pm}$ from equation (\ref{eq:Delp}). 
In other words, the momentum diffusion does not occur if $D_{p+}$ or $D_{p-}$ is zero. 
As an estimate, we may write the momentum-diffusion coefficient $\Dp$ used in the diffusion transport equation (Equation [\ref{eq:dfdt}]) as
\begin{align}
 \label{eq:Dp}
 \Dp &\approx \left< \frac{(2 \Delta p)^2}{(\Delta p)^2/D_{p+} + (\Delta p)^2/D_{p-}} \right>_\mu \nonumber \\
 &\approx \pi p^2 \left( \frac{\vw}{v} \right)^2 \frac{\left| \Omegag \right|}{\rg B^2} \int_{\left| \mu \right| > \vw/v} \dif \mu \frac{1 - \mu^2}{\left| \mu \right|} \frac{B_+^2(k_\|) B_-^2(k_\|)}{B_+^2(k_\|) + B_-^2(k_\|)} 
\end{align}
(as in the case of the homogenization of the diffusion coefficient for inhomogeneous media), where $\left< \right>_\mu$ indicates the pitch-angle averaging and $k_\|$ is the resonant wavenumber, which is a function of $\mu$ from equation (\ref{eq:kres}) \citep{Skilling1975a, Schlickeiser1989a}. 
Note that the lower end of the integration in equation (\ref{eq:Dp}) simply comes from the present approximation, equation (\ref{eq:fast}). 
Intrinsically the scattering can also occur for particles with $\left| \mu \right| < \vw/v \ll 1$ \citep[e.g.,][]{Schlickeiser1989a}. 

The integrand in equation (\ref{eq:Dp}) can be divided into two parts:
\begin{align}
 \label{eq:div}
 \frac{B_+^2(k_\|) B_-^2(k_\|)}{B_+^2(k_\|) + B_-^2(k_\|)} \propto \sum_{j=\pm} B_j^2(k_\|) - \frac{\left( \sum_{j=\pm} j B_j^2(k_\|) \right)^2}{\sum_{j=\pm} B_j^2(k_\|)} \, . 
\end{align}
Then, one can confirm that this expression corresponds to $\Dp = \left< D_{pp} - \Dmup^2/\Dmumu \right>_\mu$ given by, e.g., \citet{Schlickeiser1989a} in the approximation of equation (\ref{eq:fast}), where $D_{pp}$, $\Dmup$, and $\Dmumu$ are the Fokker--Planck coefficients (equation [\ref{eq:FP}]). 
The term $\Dmup^2/\Dmumu$ originates from particle anisotropy, which is assumed to be small in equation (\ref{eq:dfdt}), so that this term is less important compared to $D_{pp}$. 
Correspondingly, the second term in equation (\ref{eq:div}) vanishes when the cross helicity, which causes the anisotropy, is zero, $B_+^2(k_\|) = B_-^2(k_\|)$. 

Supposing zero cross helicity and zero magnetic helicity, that is, $B_+^2(k_\|) = B_-^2(k_\|) = B_\pm^2(-k_\|) (\equiv B^2(k_\|))$, we have 
\begin{align}
 \Dp &\approx \pi p^2 \left( \frac{\vw}{v} \right)^2 \frac{\left| \Omegag \right|}{\rg B^2} \int_{1/\rg} \frac{\dif k_\|}{k_\|} \left(1 - \frac{1}{(\rg k_\|)^2} \right) B^2(k_\|) \, , 
\end{align}
where the integration variable has been changed by equation (\ref{eq:kres}) \citep[see][for the influence of the helicities under the assumption of isospectral turbulence]{Dung1990b}. 
The particles keep isotropy by pitch-angle scattering on a much shorter timescale than the momentum diffusion. 
If $B^2(k_\|)$ becomes smaller for larger $\left| k_\| \right|$, the lower end of the integration is more important than the upper end. 
Then, the wavenumber $k_\| \sim 1/\rg \propto 1/p$ mainly contributes to $\Dp$.

\clearpage

\bibliography{ref}

\begin{thebibliography}{}
\expandafter\ifx\csname natexlab\endcsname\relax\def\natexlab#1{#1}\fi

\bibitem[{{Abdo} {et~al.}(2010){Abdo}, {Ackermann}, {Agudo}, {Ajello}, {Aller},
  {Aller}, {Angelakis}, {Arkharov}, {Axelsson}, {Bach}, \& et~al.}]{Abdo2010c}
{Abdo}, A.~A., {Ackermann}, M., {Agudo}, I., {et~al.} 2010, \apj, 716, 30

\bibitem[{{Abdo} {et~al.}(2011{\natexlab{a}}){Abdo}, {Ackermann}, {Ajello},
  {Baldini}, {Ballet}, {Barbiellini}, {Bastieri}, {Bechtol}, {Bellazzini},
  {Berenji}, \& et~al.}]{abdo2011a}
{Abdo}, A.~A., {Ackermann}, M., {Ajello}, M., {et~al.} 2011{\natexlab{a}},
  \apj, 736, 131

\bibitem[{{Abdo} {et~al.}(2011{\natexlab{b}}){Abdo}, {Ackermann}, {Ajello},
  {Allafort}, {Baldini}, {Ballet}, {Barbiellini}, {Baring}, {Bastieri},
  {Bechtol}, \& et~al.}]{abdo2011b}
---. 2011{\natexlab{b}}, \apj, 727, 129

\bibitem[{{Aharonian} {et~al.}(2009){Aharonian}, {Akhperjanian}, {Anton},
  {Barres de Almeida}, {Bazer-Bachi}, {Becherini}, {Behera}, {Bernl{\"o}hr},
  {Boisson}, {Bochow}, \& et~al.}]{Aharonian2009a}
{Aharonian}, F., {Akhperjanian}, A.~G., {Anton}, G., {et~al.} 2009, \apjl, 696,
  L150

\bibitem[{{Asano} \& {Hayashida}(2015)}]{Asano2015a}
{Asano}, K., \& {Hayashida}, M. 2015, \apjl, 808, L18

\bibitem[{{Asano} {et~al.}(2014){Asano}, {Takahara}, {Kusunose}, {Toma}, \&
  {Kakuwa}}]{Asano2014a}
{Asano}, K., {Takahara}, F., {Kusunose}, M., {Toma}, K., \& {Kakuwa}, J. 2014,
  \apj, 780, 64

\bibitem[{{Beckmann} \& {Shrader}(2012)}]{Beckmann2012a}
{Beckmann}, V., \& {Shrader}, C.~R. 2012, {Active Galactic Nuclei, Weinheim:
  Wiley-VCH}

\bibitem[{{Blandford} \& {Eichler}(1987)}]{Blandford1987a}
{Blandford}, R., \& {Eichler}, D. 1987, \physrep, 154, 1

\bibitem[{{Blasi}(2013)}]{Blasi2013a}
{Blasi}, P. 2013, \aapr, 21, 70

\bibitem[{{Blumenthal} \& {Gould}(1970)}]{Blumenthal1970a}
{Blumenthal}, G.~R., \& {Gould}, R.~J. 1970, Rev. Mod. Phys., 42, 237

\bibitem[{{Bogdan} \& {Schlickeiser}(1985)}]{Bogdan1985a}
{Bogdan}, T.~G., \& {Schlickeiser}, R. 1985, \aap, 143, 23

\bibitem[{{Borovsky} \& {Eilek}(1986)}]{Borovsky1986a}
{Borovsky}, J.~E., \& {Eilek}, J.~A. 1986, \apj, 308, 929

\bibitem[{{Brunetti} \& {Lazarian}(2007)}]{Brunetti2007a}
{Brunetti}, G., \& {Lazarian}, A. 2007, \mnras, 378, 245

\bibitem[{{Dermer} \& {Lott}(2012)}]{Dermer2012a}
{Dermer}, C., \& {Lott}, B. 2012, J. Phys. Conf. Ser., 355, 012010

\bibitem[{{Dermer} {et~al.}(1996){Dermer}, {Miller}, \& {Li}}]{Dermer1996a}
{Dermer}, C.~D., {Miller}, J.~A., \& {Li}, H. 1996, \apj, 456, 106

\bibitem[{{Dobrowolny} {et~al.}(1980){Dobrowolny}, {Mangeney}, \&
  {Veltri}}]{Dobrowolny1980a}
{Dobrowolny}, M., {Mangeney}, A., \& {Veltri}, P. 1980, Physical Review
  Letters, 45, 144

\bibitem[{{Drury}(1987)}]{Drury1987a}
{Drury}, L.~O. 1987, Irish Astronomical Journal, 18, 28

\bibitem[{{Dung} \& {Schlickeiser}(1990)}]{Dung1990b}
{Dung}, R., \& {Schlickeiser}, R. 1990, \aap, 240, 537

\bibitem[{{Eichler}(1979)}]{Eichler1979b}
{Eichler}, D. 1979, \apj, 229, 413

\bibitem[{{Eilek}(1979)}]{Eilek1979a}
{Eilek}, J.~A. 1979, \apj, 230, 373

\bibitem[{{Eilek} \& {Henriksen}(1984)}]{Eilek1984a}
{Eilek}, J.~A., \& {Henriksen}, R.~N. 1984, \apj, 277, 820

\bibitem[{{Fermi}(1949)}]{Fermi1949a}
{Fermi}, E. 1949, Physical Review, 75, 1169

\bibitem[{{Finke} {et~al.}(2008){Finke}, {Dermer}, \&
  {B{\"o}ttcher}}]{Finke2008a}
{Finke}, J.~D., {Dermer}, C.~D., \& {B{\"o}ttcher}, M. 2008, \apj, 686, 181

\bibitem[{{Garson} {et~al.}(2010){Garson}, {Baring}, \&
  {Krawczynski}}]{Garson2010a}
{Garson}, III, A.~B., {Baring}, M.~G., \& {Krawczynski}, H. 2010, \apj, 722,
  358

\bibitem[{{Giebels} {et~al.}(2007){Giebels}, {Dubus}, \&
  {Kh{\'e}lifi}}]{Giebels2007a}
{Giebels}, B., {Dubus}, G., \& {Kh{\'e}lifi}, B. 2007, \aap, 462, 29

\bibitem[{{Goldstein} {et~al.}(1995){Goldstein}, {Roberts}, \&
  {Matthaeus}}]{Goldstein1995a}
{Goldstein}, M.~L., {Roberts}, D.~A., \& {Matthaeus}, W.~H. 1995, \araa, 33,
  283

\bibitem[{{Hall} \& {Sturrock}(1967)}]{Hall1967a}
{Hall}, D.~E., \& {Sturrock}, P.~A. 1967, Phys. Fluids, 10, 2620

\bibitem[{{Inoue} \& {Takahara}(1996)}]{Inoue1996a}
{Inoue}, S., \& {Takahara}, F. 1996, \apj, 463, 555

\bibitem[{{Jones}(1968)}]{Jones1968a}
{Jones}, F.~C. 1968, Physical Review, 167, 1159

\bibitem[{{Kakuwa} {et~al.}(2015){Kakuwa}, {Toma}, {Asano}, {Kusunose}, \&
  {Takahara}}]{Kakuwa2015a}
{Kakuwa}, J., {Toma}, K., {Asano}, K., {Kusunose}, M., \& {Takahara}, F. 2015,
  \mnras, 449, 551

\bibitem[{{Kataoka} {et~al.}(1999){Kataoka}, {Mattox}, {Quinn}, {Kubo},
  {Makino}, {Takahashi}, {Inoue}, {Hartman}, {Madejski}, {Sreekumar}, \&
  {Wagner}}]{Kataoka1999a}
{Kataoka}, J., {Mattox}, J.~R., {Quinn}, J., {et~al.} 1999, \apj, 514, 138

\bibitem[{{Katarzy{\'n}ski} {et~al.}(2006{\natexlab{a}}){Katarzy{\'n}ski},
  {Ghisellini}, {Mastichiadis}, {Tavecchio}, \& {Maraschi}}]{Katarzynski2006b}
{Katarzy{\'n}ski}, K., {Ghisellini}, G., {Mastichiadis}, A., {Tavecchio}, F.,
  \& {Maraschi}, L. 2006{\natexlab{a}}, \aap, 453, 47

\bibitem[{{Katarzy{\'n}ski} {et~al.}(2006{\natexlab{b}}){Katarzy{\'n}ski},
  {Ghisellini}, {Tavecchio}, {Gracia}, \& {Maraschi}}]{Katarzynski2006a}
{Katarzy{\'n}ski}, K., {Ghisellini}, G., {Tavecchio}, F., {Gracia}, J., \&
  {Maraschi}, L. 2006{\natexlab{b}}, \mnras, 368, L52

\bibitem[{{Kimura} {et~al.}(2015){Kimura}, {Murase}, \& {Toma}}]{Kimura2015a}
{Kimura}, S.~S., {Murase}, K., \& {Toma}, K. 2015, \apj, 806, 159

\bibitem[{{Kino} {et~al.}(2002){Kino}, {Takahara}, \& {Kusunose}}]{Kino2002a}
{Kino}, M., {Takahara}, F., \& {Kusunose}, M. 2002, \apj, 564, 97

\bibitem[{{Krawczynski} {et~al.}(2004){Krawczynski}, {Hughes}, {Horan},
  {Aharonian}, {Aller}, {Aller}, {Boltwood}, {Buckley}, {Coppi}, {Fossati},
  {G{\"o}tting}, {Holder}, {Horns}, {Kurtanidze}, {Marscher}, {Nikolashvili},
  {Remillard}, {Sadun}, \& {Schr{\"o}der}}]{Krawczynski2004a}
{Krawczynski}, H., {Hughes}, S.~B., {Horan}, D., {et~al.} 2004, \apj, 601, 151

\bibitem[{{Kulsrud}(2005)}]{Kulsrud2005a}
{Kulsrud}, R.~M. 2005, {Plasma physics for astrophysics, Princeton Univ. Press,
  Princeton}

\bibitem[{{Kusunose} \& {Takahara}(2008)}]{Kusunose2008a}
{Kusunose}, M., \& {Takahara}, F. 2008, \apj, 682, 784

\bibitem[{{Lefa} {et~al.}(2011{\natexlab{a}}){Lefa}, {Aharonian}, \&
  {Rieger}}]{Lefa2011b}
{Lefa}, E., {Aharonian}, F.~A., \& {Rieger}, F.~M. 2011{\natexlab{a}}, \apjl,
  743, L19

\bibitem[{{Lefa} {et~al.}(2011{\natexlab{b}}){Lefa}, {Rieger}, \&
  {Aharonian}}]{Lefa2011a}
{Lefa}, E., {Rieger}, F.~M., \& {Aharonian}, F. 2011{\natexlab{b}}, \apj, 740,
  64

\bibitem[{{Li} \& {Kusunose}(2000)}]{Li2000a}
{Li}, H., \& {Kusunose}, M. 2000, \apj, 536, 729

\bibitem[{{Longair}(1992)}]{Longair1992a}
{Longair}, M.~S. 1992, {High Energy Astrophysics. Vol.1: Particles, photons and
  their detection, Cambridge Univ. Press, Cambridge}

\bibitem[{{Maraschi} {et~al.}(1994){Maraschi}, {Ghisellini}, \&
  {Celotti}}]{Maraschi1994a}
{Maraschi}, L., {Ghisellini}, G., \& {Celotti}, A. 1994, in Proc. IAU Symp.,
  Vol. 159, Multi-Wavelength Continuum Emission of AGN, ed. T.~{Courvoisier} \&
  A.~{Blecha}, 221--232

\bibitem[{{Marscher}(2009)}]{Marscher2009a}
{Marscher}, A.~P. 2009, ArXiv e-prints, arXiv:0909.2576

\bibitem[{{Miller} {et~al.}(1996){Miller}, {Larosa}, \& {Moore}}]{Miller1996a}
{Miller}, J.~A., {Larosa}, T.~N., \& {Moore}, R.~L. 1996, \apj, 461, 445

\bibitem[{{Miller} \& {Roberts}(1995)}]{Miller1995a}
{Miller}, J.~A., \& {Roberts}, D.~A. 1995, \apj, 452, 912

\bibitem[{{Ohira}(2013)}]{Ohira2013a}
{Ohira}, Y. 2013, \apjl, 767, L16

\bibitem[{{Ohno} {et~al.}(2002){Ohno}, {Takizawa}, \& {Shibata}}]{Ohno2002a}
{Ohno}, H., {Takizawa}, M., \& {Shibata}, S. 2002, \apj, 577, 658

\bibitem[{{Ostrowski} \& {Siemieniec-Ozi{\c e}b{\l}o}(1997)}]{Ostrowski1997a}
{Ostrowski}, M., \& {Siemieniec-Ozi{\c e}b{\l}o}, G. 1997, Astropart. Phys., 6,
  271

\bibitem[{{Parker} \& {Tidman}(1958)}]{Parker1958a}
{Parker}, E.~N., \& {Tidman}, D.~A. 1958, Physical Review, 111, 1206

\bibitem[{{Pohl} \& {Schlickeiser}(2000)}]{Pohl2000a}
{Pohl}, M., \& {Schlickeiser}, R. 2000, \aap, 354, 395

\bibitem[{{Ptuskin}(1988)}]{Ptuskin1988a}
{Ptuskin}, V.~S. 1988, Soviet Astronomy Letters, 14, 255

\bibitem[{{Rani} {et~al.}(2013){Rani}, {Krichbaum}, {Fuhrmann}, {B{\"o}ttcher},
  {Lott}, {Aller}, {Aller}, {Angelakis}, {Bach}, {Bastieri}, {Falcone},
  {Fukazawa}, {Gabanyi}, {Gupta}, {Gurwell}, {Itoh}, {Kawabata}, {Krips},
  {L{\"a}hteenm{\"a}ki}, {Liu}, {Marchili}, {Max-Moerbeck}, {Nestoras},
  {Nieppola}, {Quintana-Lacaci}, {Readhead}, {Richards}, {Sasada}, {Sievers},
  {Sokolovsky}, {Stroh}, {Tammi}, {Tornikoski}, {Uemura}, {Ungerechts},
  {Urano}, \& {Zensus}}]{Rani2013a}
{Rani}, B., {Krichbaum}, T.~P., {Fuhrmann}, L., {et~al.} 2013, \aap, 552, A11

\bibitem[{{Rybicki} \& {Lightman}(1979)}]{Rybicki1979a}
{Rybicki}, G.~B., \& {Lightman}, A.~P. 1979, {Radiative Processes in
  Astrophysics. Wiley, New York}

\bibitem[{{Saug{\'e}} \& {Henri}(2004)}]{Sauge2004a}
{Saug{\'e}}, L., \& {Henri}, G. 2004, \apj, 616, 136

\bibitem[{{Schlickeiser}(1984)}]{Schlickeiser1984a}
{Schlickeiser}, R. 1984, \aap, 136, 227

\bibitem[{{Schlickeiser}(1985)}]{Schlickeiser1985a}
---. 1985, \aap, 143, 431

\bibitem[{{Schlickeiser}(1989)}]{Schlickeiser1989a}
---. 1989, \apj, 336, 243

\bibitem[{{Schlickeiser}(2002)}]{Schlickeiser2002a}
---. 2002, {Cosmic Ray Astrophysics. Springer-Verlag, Berlin}

\bibitem[{{Schlickeiser} \& {Dermer}(2000)}]{Schlickeiser2000a}
{Schlickeiser}, R., \& {Dermer}, C.~D. 2000, \aap, 360, 789

\bibitem[{{Schlickeiser} \& {Miller}(1998)}]{Schlickeiser1998a}
{Schlickeiser}, R., \& {Miller}, J.~A. 1998, \apj, 492, 352

\bibitem[{{Schlickeiser} {et~al.}(2002){Schlickeiser}, {Vainio},
  {B{\"o}ttcher}, {Lerche}, {Pohl}, \& {Schuster}}]{Schlickeiser2002b}
{Schlickeiser}, R., {Vainio}, R., {B{\"o}ttcher}, M., {et~al.} 2002, \aap, 393,
  69

\bibitem[{{Sikora} \& {Madejski}(2001)}]{Sikora2001a}
{Sikora}, M., \& {Madejski}, G. 2001, in AIP Conf. Proc., Vol. 558, High Energy
  Gamma-Ray Astronomy. Am. Inst. Phys., New York, ed. F.~A. {Aharonian} \&
  H.~J. {V{\"o}lk}, 275

\bibitem[{{Skilling}(1975)}]{Skilling1975a}
{Skilling}, J. 1975, \mnras, 172, 557

\bibitem[{{Stawarz} \& {Petrosian}(2008)}]{Stawarz2008a}
{Stawarz}, {\L}., \& {Petrosian}, V. 2008, \apj, 681, 1725

\bibitem[{{Tagliaferri} {et~al.}(2008){Tagliaferri}, {Foschini}, {Ghisellini},
  {Maraschi}, {Tosti}, {Albert}, {Aliu}, {Anderhub}, {Antoranz}, {Baixeras},
  {Barrio}, {Bartko}, {Bastieri}, {Becker}, {Bednarek}, {Bedyugin}, {Berger},
  {Bigongiari}, {Biland}, {Bock}, {Bordas}, {Bosch-Ramon}, {Bretz},
  {Britvitch}, {Camara}, {Carmona}, {Chilingarian}, {Coarasa}, {Commichau},
  {Contreras}, {Cortina}, {Costado}, {Curtef}, {Danielyan}, {Dazzi}, {De
  Angelis}, {Delgado}, {de los Reyes}, {De Lotto}, {Dorner}, {Doro}, {Errando},
  {Fagiolini}, {Ferenc}, {Fern{\'a}ndez}, {Firpo}, {Fonseca}, {Font}, {Fuchs},
  {Galante}, {Garc{\'{\i}}a-L{\'o}pez}, {Garczarczyk}, {Gaug}, {Giller},
  {Goebel}, {Hakobyan}, {Hayashida}, {Hengstebeck}, {Herrero}, {H{\"o}hne},
  {Hose}, {Huber}, {Hsu}, {Jacon}, {Jogler}, {Kosyra}, {Kranich}, {Kritzer},
  {Laille}, {Lindfors}, {Lombardi}, {Longo}, {L{\'o}pez}, {Lorenz}, {Majumdar},
  {Maneva}, {Mannheim}, {Mariotti}, {Mart{\'{\i}}nez}, {Mazin}, {Merck},
  {Meucci}, {Meyer}, {Miranda}, {Mirzoyan}, {Mizobuchi}, {Moralejo}, {Nieto},
  {Nilsson}, {Ninkovic}, {O{\~n}a-Wilhelmi}, {Otte}, {Oya}, {Panniello},
  {Paoletti}, {Paredes}, {Pasanen}, {Pascoli}, {Pauss}, {Pegna}, {Persic},
  {Peruzzo}, {Piccioli}, {Prandini}, {Puchades}, {Raymers}, {Rhode},
  {Rib{\'o}}, {Rico}, {Rissi}, {Robert}, {R{\"u}gamer}, {Saggion}, {Saito},
  {S{\'a}nchez}, {Sartori}, {Scalzotto}, {Scapin}, {Schmitt}, {Schweizer},
  {Shayduk}, {Shinozaki}, {Shore}, {Sidro}, {Sillanp{\"a}{\"a}}, {Sobczynska},
  {Spanier}, {Stamerra}, {Stark}, {Takalo}, {Tavecchio}, {Temnikov}, {Tescaro},
  {Teshima}, {Torres}, {Turini}, {Vankov}, {Venturini}, {Vitale}, {Wagner},
  {Wibig}, {Wittek}, {Zandanel}, {Zanin}, {Zapatero}, \& {MAGIC
  Collaboration}}]{Tagliaferri2008a}
{Tagliaferri}, G., {Foschini}, L., {Ghisellini}, G., {et~al.} 2008, \apj, 679,
  1029

\bibitem[{{Tavecchio} {et~al.}(2010){Tavecchio}, {Ghisellini}, {Ghirlanda},
  {Foschini}, \& {Maraschi}}]{Tavecchio2010a}
{Tavecchio}, F., {Ghisellini}, G., {Ghirlanda}, G., {Foschini}, L., \&
  {Maraschi}, L. 2010, \mnras, 401, 1570

\bibitem[{{Terasawa}(1989)}]{Terasawa1989a}
{Terasawa}, T. 1989, Washington DC American Geophysical Union Geophysical
  Monograph Series, 53, 41

\bibitem[{{Tramacere} {et~al.}(2011){Tramacere}, {Massaro}, \&
  {Taylor}}]{Tramacere2011a}
{Tramacere}, A., {Massaro}, E., \& {Taylor}, A.~M. 2011, \apj, 739, 66

\bibitem[{{Urry} \& {Padovani}(1995)}]{Urry1995a}
{Urry}, C.~M., \& {Padovani}, P. 1995, \pasp, 107, 803

\bibitem[{{Ushio} {et~al.}(2009){Ushio}, {Tanaka}, {Madejski}, {Takahashi},
  {Hayashida}, {Kataoka}, {Mazin}, {R{\"u}gamer}, {Sato}, {Teshima}, {Wagner},
  \& {Yaji}}]{Ushio2009a}
{Ushio}, M., {Tanaka}, T., {Madejski}, G., {et~al.} 2009, \apj, 699, 1964

\bibitem[{{Ushio} {et~al.}(2010){Ushio}, {Stawarz}, {Takahashi}, {Paneque},
  {Madejski}, {Hayashida}, {Kataoka}, {Tanaka}, {Tanaka}, \&
  {Ostrowski}}]{Ushio2010a}
{Ushio}, M., {Stawarz}, {\L}., {Takahashi}, T., {et~al.} 2010, \apj, 724, 1509

\bibitem[{{Wentzel}(1974)}]{Wentzel1974a}
{Wentzel}, D.~G. 1974, \araa, 12, 71

\bibitem[{{Zhang} {et~al.}(2012){Zhang}, {Liang}, {Zhang}, \&
  {Bai}}]{Zhang2012a}
{Zhang}, J., {Liang}, E.-W., {Zhang}, S.-N., \& {Bai}, J.~M. 2012, \apj, 752,
  157

\bibitem[{{Zheng} \& {Zhang}(2011)}]{Zheng2011a}
{Zheng}, Y.~G., \& {Zhang}, L. 2011, \apj, 728, 105

\bibitem[{{Zhou} \& {Matthaeus}(1990)}]{Zhou1990a}
{Zhou}, Y., \& {Matthaeus}, W.~H. 1990, \jgr, 95, 14881

\bibitem[{{Zhou} {et~al.}(2004){Zhou}, {Matthaeus}, \& {Dmitruk}}]{Zhou2004a}
{Zhou}, Y., {Matthaeus}, W.~H., \& {Dmitruk}, P. 2004, Reviews of Modern
  Physics, 76, 1015

\end{thebibliography}

\begin{table*}
\begin{center}
\caption{Efficiency of the damping of turbulence by electron acceleration, $\etadam(\kres)$ (equation [\ref{eq:etadam2}]), for some previous works at the wavenumber $\kres = \kres(\gamma)$. 
The damping efficiency higher than unity, $\etadam(k) > 1$, implies that the wavenumber spectrum of the turbulent fields is being affected by the electron acceleration at wavenumber $k$. 
Physical quantities used for the evaluation are also shown. 
See also footnote \ref{foo:katarzynski2006b}. }
\label{tab:etadam}
\begin{tabular}{llllllll}
\tableline\tableline
             & Refs          & $\gamma$         & $\gamma^3 \nele(\gamma)$ & $\etasyn(\gamma)$ & $B$ & $\zeta(\kres)$ & $\etadam(\kres)$ \\
Unit         &               &                  & (cm$^{-3}$)              &                   & (G) &                &           \\
\tableline
1ES 0229+200 & 1             & $1.5\times 10^5$ & $8.4\times 10^{11}$ & 1    & 0.07  & $\lesssim 0.1$      & $\gtrsim 40$    \\
Mrk 501      & 2$^{{\rm a}}$ & $5.0\times 10^6$ & $1.0\times 10^{14}$ & 1.7  & 0.05  & $\lesssim 0.1$      & $\gtrsim 6000$  \\
Mrk 501      & 2$^{{\rm b}}$ & $3.4\times 10^6$ & $1.5\times 10^{12}$ & 1.8  & 0.11  & $\lesssim 0.1$      & $\gtrsim 30$    \\
Mrk 501      & 3             & $1.0\times 10^6$ & $1.6\times 10^{8}$  & 1    & 0.014 & $1.9\times 10^{-5}$ & 500             \\
Mrk 421      & 3             & $1.5\times 10^5$ & $6.1\times 10^{8}$  & 1    & 0.081 & $5.6\times 10^{-6}$ & 400             \\
Mrk 421      & 4             & $3.0\times 10^5$ & $1.3\times 10^{8}$  & 1    & 0.043 & $\lesssim 0.1$      & $\gtrsim 0.01$  \\
1ES 1101-232 & 4             & $7.0\times 10^5$ & $1.7\times 10^{8}$  & 0.3  & 0.015 & $\lesssim 0.1$      & $\gtrsim 0.1$   \\
\tableline
\end{tabular}
\tablerefs{(1) \citet[Figure 10]{Lefa2011a}; (2) \citet[Figures 3$^{{\rm a}}$ and 4$^{{\rm b}}$]{Katarzynski2006b}; (3) \citet[Figure 2]{Kakuwa2015a}; (4) \citet[Figures 5 and 6]{Asano2014a}}
\end{center}
\end{table*}

\end{document}